\documentclass[pre,amsmath,amssymb,aps,twocolumn]{revtex4-2}
\usepackage{graphicx}
\usepackage{color}
\usepackage{float}
\usepackage{times}

\usepackage[colorlinks=true,linkcolor=blue,urlcolor=blue,citecolor=blue]{hyperref}

\usepackage{tikz}

\usetikzlibrary{arrows,shapes,positioning,decorations.pathmorphing}
\usetikzlibrary{decorations.markings}
\usetikzlibrary{snakes}
\tikzset{snake it/.style={decorate, decoration=snake,segment length=3mm}}
\tikzstyle arrowstyle=[scale=1]
\tikzstyle directed=[postaction={decorate,decoration={markings,
    mark=at position .65 with {\arrow[arrowstyle]{stealth}}}}]
\tikzstyle endreversedirected=[postaction={decorate,decoration={markings,
    mark=at position 1.0 with {\arrow[arrowstyle]{stealth}}}}]
\tikzstyle enddirected=[postaction={decorate,decoration={markings,
    mark=at position 1.0 with {\arrow[arrowstyle]{stealth}}}}]
\tikzstyle reverse directed=[postaction={decorate,decoration={markings,
    mark=at position .65 with {\arrowreversed[arrowstyle]{stealth};}}}]

\newcommand{\Mathematica}[1]{}
\newcommand{\Eq}[1]{Eq.~(\ref{#1})}
\newcommand{\eq}[1]{(\ref{#1})}

\newcommand{\half}{\frac12}

\newcommand{\bea}{\begin{eqnarray}}
\newcommand{\eea}{\end{eqnarray}}
\newcommand{\beq}{\begin{equation}}
\newcommand{\eeq}{\end{equation}}

\newcommand{\be}{\begin{equation}}
\newcommand{\ee}{\end{equation}}

\newcommand{\rme}{\mathrm{e}}
\newcommand{\rmd}{\mathrm{d}}
\newcommand{\nn}{\nonumber}
\renewcommand{\epsilon}{\varepsilon}
\newcommand{\nott}[1]{}
\newcommand{\Fig}[1]{\includegraphics[width=\columnwidth]{./#1}} 
\newcommand{\fig}[2]{\includegraphics[width=#1\columnwidth]{./#2}}

\newlength{\bilderlength}

\newcommand{\1}{1\hspace*{-0.5ex}{\rm l}}

\graphicspath{{figures/},{}}

\renewcommand{\log}{\ln}

\renewcommand{\paragraph}{\subsubsection*}

\begin{document}

\title{Universal Force Correlations in an RNA-DNA Unzipping Experiment}
\author{Kay J\"org Wiese${}^{1}$, Mathilde Bercy${}^{2}$, Lena Melkonyan${}^{2}$, Thierry Bizebard${}^{3}$}

\affiliation{${}^{1}$\mbox{Laboratoire de Physique de l'\'Ecole Normale Sup\'erieure, ENS, Universit\'e PSL, CNRS, Sorbonne} 
\mbox{Universit\'e, Universit\'e Paris-Diderot, Sorbonne Paris Cit\'e, 24 rue Lhomond, 75005 Paris, France.}\\
${}^{2}$UMR 7615, Ecole Sup\'erieure de Physique et Chimie Industrielles de la Ville de Paris, Paris, France.\\
${}^{3}$UMR  8104, Institut Cochin, 22 rue M\'echain, 75014 Paris, France.}

\begin{abstract}
We study unzipping of a complementary RNA-DNA helix applied to an external force, focusing on the force-force correlations. 
 While at the microscopic level these are given by the sequence, the experiment measures   effective, macroscopic correlations. The latter are sequence-independent, i.e.\ universal, and constitute the central object of the underlying field theory of disordered systems.  
 Comparing field-theory predictions   and the exact solution of a 1-d toy model with the experiment, we find an excellent agreement, confirming   fundamental theoretical concepts via a biologically inspired experiments. 
\end{abstract}

\maketitle

\section{Introduction}
\begin{figure}[b]
{\fboxsep0mm\mbox{\begin{tikzpicture}
\node at (4.3,1.5) {\includegraphics[width=7.5cm]{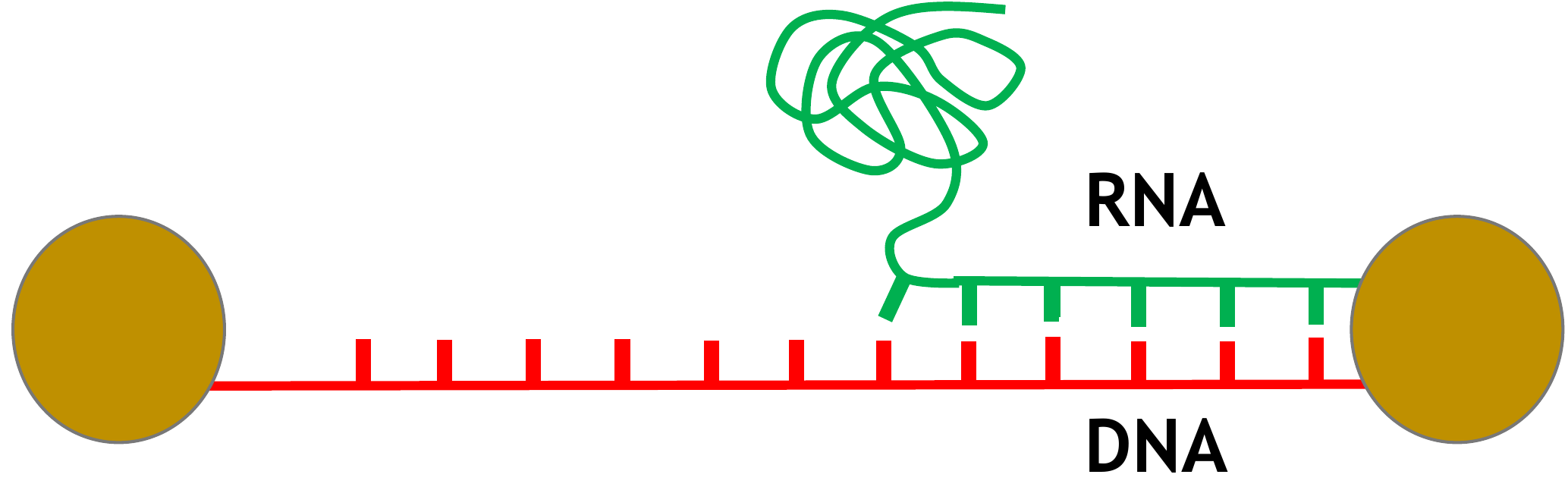}};
\draw [blue,thick] (1,0) parabola (2,1.5) ;
\draw [blue,thick] (1,0) parabola (0,1.5) ;
\draw [blue,thick] (7.6,0) parabola (8.6,1.5) ;
\draw [blue,thick] (7.6,0) parabola (6.6,1.5) ;
\draw [thick,dashed] (1,0) -- (1,-0.35);
\draw [thick,dashed] (7.6,0) -- (7.6,-0.35);
\draw [thick,->] (4,-0.25) -- (1,-0.25) ;
\draw [thick,->] (4.6,-0.25) -- (7.6,-0.25) ;
\node at (4.3,-0.25) {$w$};
\draw [thick,dashed]  (1.1,3.05) -- (1.1,1);
\draw [thick,dashed] (7.5,3.05) -- (7.5,1) ;
\draw [thick,->] (4,3) -- (1.1,3) ;
\draw [thick,->] (4.6,3) -- (7.5,3) ;
\node at (4.3,3) {$u$};
\end{tikzpicture}}}
\caption{Peeling of a RNA-DNA double strand. The RNA sequence  is   from subunit 23S of the ribosome in E.~Coli, prolonged to attach the beads (with a much larger radius than drawn here). The DNA sequence is its complement.  Drawing not to scale.}
\label{f:peeling}
\end{figure}
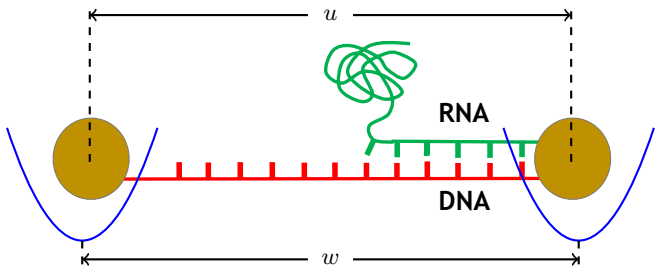

\begin{figure}[b]
\centerline{\fboxsep0mm
{\setlength{\unitlength}{1cm}\begin{picture}(8.65,5.75)
\put(0,0){\fig{1}{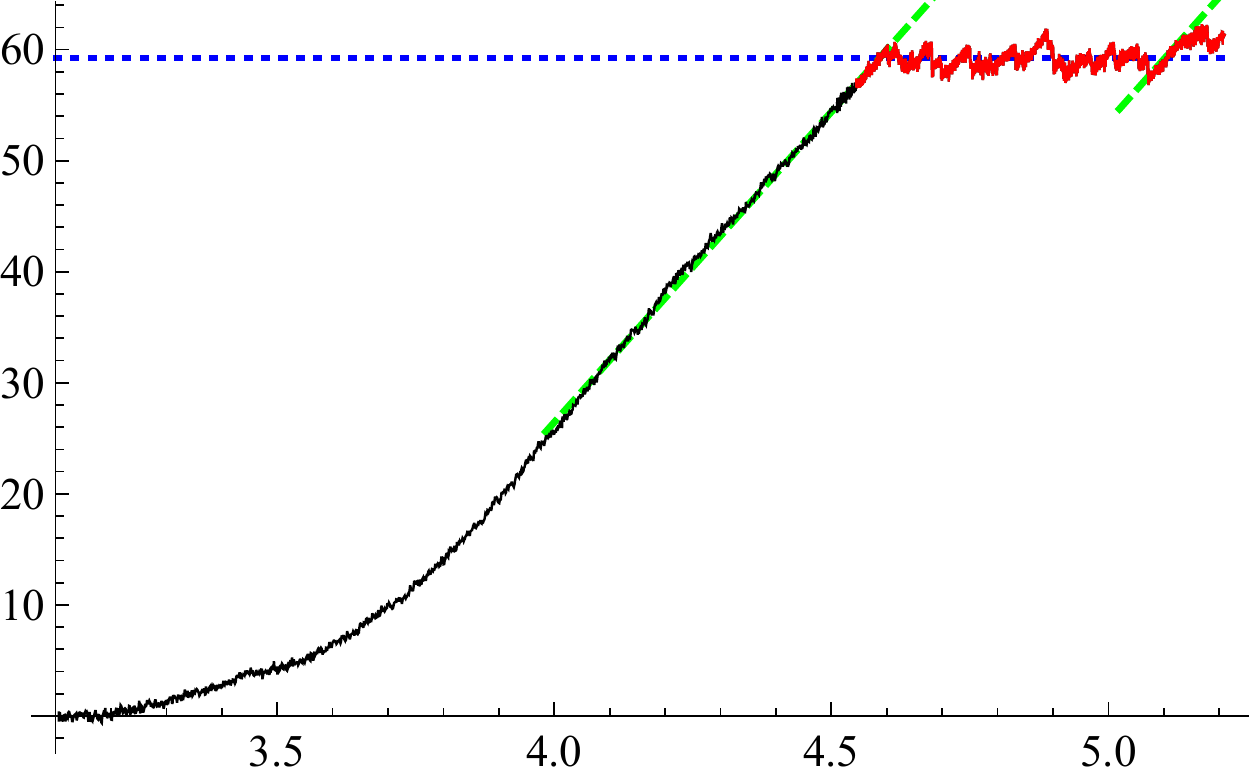}}
\put(7.8,0.6){$w[\mu\rm m]$}
\put(0,5.5){$F[\rm pN]$}
\end{picture}}}
\caption{A sample force-extension curve. For the data analysis we only use the last part of the curve, the plateau (in red).
On this plateau, the force fluctuates around its critical value of about $60 \rm pN$. 
The extension $w$ starts at $3\mu\rm m$, which is the sum of the unstretched molecule plus twice the radius of the beads ($2 \times 1 \mu m$). The effective stiffness  $m^{2}$ in \Eq{1} is estimated from the slope  of the green dashed lines as $m^{2} = 55\pm5  \rm pN/\mu m$  at the beginning of the plateau, which remains at least approximately  the same  at the end of the plateau.
The driving velocity is about    $7 \rm n m/s$, corresponding to   $42 \,\rm \mbox{nucleotides}/s$ as in the cell \cite{GottaMillerFrench1991}. }
\label{f:example}
\end{figure}

The  amount of biological data is growing steadily, reaching about 
$2.5 \times 10^{16}$ Bytes in 2015 \cite{StephensLeeFaghriCampbellZhaiEfronIyerSchatzSinhaRobinson2015}, on par  with   astronomy, youtube and Twitter. 
An important question is what can be learned from these data, and what cannot? Depending on their specialisation,   scientists usually ask different, and seemingly unrelated questions. Here we study unzipping in the peeling mode \cite{unzipping} of a complementary RNA-DNA double strand, using a sequence obtained from ribosomal RNA. As shown on Fig.~\ref{f:peeling}, at one end the double helix is attached with its both strands to a bead, whereas on the other end only the DNA-strand is. Pulling on the beads with an optical tweezer \cite{Ashkin1970} the RNA strand peels off. What is measured is the force-extension curve, of which an example is given on Fig.~\ref{f:example}.

Rather complementary questions can now be asked: 
\begin{enumerate}
\item[(i)] What can one learn about the specific {\em biological system}?
\item[(ii)] Are there  observables which are independent of the chosen nucleotide sequence, thus {\em universal}? 
\item[(iii)] How does understanding the {\em universal signal} help  to analyse the  {\em biological system}? What limitations does it impose?
\end{enumerate}
The first question is at the origin and design  of the experiment \cite{BercyThesis,MelkonyanThesis,MelkonyanBercyBizebardBockelmann2019}, 
which aims at understanding assembly  of the large subunit of the ribosome,   and where all experimental parameters can be found.
Here we address the second question, aiming at understanding its universal, sequence-independent properties. This  allows us to    address the third question, helping to better extract   biologically relevant data.

Consider the force-extension curve on Fig.~\ref{f:example}. Applying no force,  the RNA-DNA double strand is in an equilibrated  coiled state, with its end-to-end distance being roughly $0.8\mu m$. Since the beads are sitting in an optical trap, their distance, or more specifically the distance  $w$ between the two minima of the trap,  is the control parameter. Increasing $w$, the RNA-DNA double strand is   stretched,  reflected in an increase in the measured force $F$.
Finally part of the RNA sequence   peels off \cite{LegerRomanoSarkarRobertBourdieuChatenayMarko1999}, leading to a first   drop in the force-extension curve.
Increasing $w$  further    leads to   more  force drops resulting in an {\em almost}  constant force. This {\em plateau regime} is marked in red on Fig.~\ref{f:example}.   Increasing $w$ further,    peeling can no longer reduce the force, and the latter   increases again, eventually leading to the breakage of the DNA molecule (not shown).  
If instead of  $w$ the applied force $F$ were controled, as in experiments with magnetic tweezers \cite{StrickAllemandBensimonBensimonCroquette1996,DeVlaminckDekker2012}, a phase transition at $F_{\rm c}$ could be observed between a closed and an open state \cite{Nelson2005}.

The aim of this letter is to analyse the force fluctuations on the plateau, i.e.\ the saw-tooth shaped signal on top of the critical force. This kind of signal is frequent in nature, and at the heart of the so-called {\em depinning transition}: It arises in a plethora of situations: Barkhausen noise in magnets \cite{Barkhausen1919,SethnaDahmenMyers2001} (audible as the rustle in old-style telephones), depinning of a contactline \cite{LeDoussalWieseMoulinetRolley2009} (the line where coffee and air meet in a cup, or in drops on a windshield), earthquakes \cite{GutenbergRichter1956}, vortices in high-temperature superconductors \cite{BlatterFeigelmanGeshkenbeinLarkinVinokur1994},  to name a few. The largest such system on earth is the movement of tectonic plates in the outer crust of the earth, where  the resulting force drops are earthquakes. The smallest system the authors are aware of is the peeling experiment studied here.  Yet,  all these systems have a very similar phenomenology: In each case, a control parameter $w$ is increased, leading to an increase in tension of the elastic object,  released via a succession of force drops. Being omnipresent, many theoretical models and mechanisms have been proposed for this {\em depinning transition}, starting from the chaos induced in the Burridge-Knopoff model of 1967 \cite{BurridgeKnopoff1967}, over toy models for magnets \cite{AlessandroBeatriceBertottiMontorsi1990,AlessandroBeatriceBertottiMontorsi1990b}, to sophisticated field theoretic work using functional RG \cite{NattermannStepanowTangLeschhorn1992,NarayanDSFisher1993a,BucheliWagnerGeshkenbeinLarkinBlatter1998,ChauveLeDoussalWiese2000a,LeDoussalWieseChauve2002,LeDoussalWieseChauve2003}. Today it is   understood that the minimal ingredients are
\begin{enumerate}
\item[(i)] a random force (the {\em disorder}),
\item[(ii)] an elastic coupling to an external control parameter,
\item[(iii)] an overdamped dynamics.
\end{enumerate}
In the experiment considered here, the random force comes from the {\em seemingly} random RNA sequence of the ribosome \cite{footnote2}. 
The elastic coupling to an external control parameter is given by the beads attached to the ends of the strands sitting in the harmonic traps with distance $w$.
Finally, an overdamped  dynamics is typical for small systems immersed into a solvent,   where inertia plays a negligible role.

\section{Theory}
The measured force can be expressed as \cite{footnote-stiffness-trap}
\begin{equation}\label{1}
F = m^2 (w-u) \ , 
\end{equation}
where $w$ is the   distance between the two traps, and $u$ the distance between the two beads, see Fig.~\ref{f:peeling}.
 This corresponds to an energy ${\cal E}= \frac{m^2}{2} (w-u)^2 $
where $m^2$ is the combined strength of the traps,    the elasticity of the partially unzipped double strand, and the attached single strand. What is measured in the experiment is the force given in Eq.~\eq{1}. 
An example   is shown on Fig.~\ref{f:example}. Upon increasing the trap distance $w$, the force first
increases until it reaches a plateau regime where it is almost constant,  $\left< F \right> \approx F_{\rm c}= 60 \rm pN$. 
We are interested in  the correlations of the force fluctuations in the plateau regime (red  on Fig.~\ref{f:example}),  i.e.\ the connected expectations
\begin{eqnarray}
\Delta(w,w') &:=& \left< F(w) F(w') \right>^{\rm c}  \\
&\equiv& \left< \big[F(w)-\langle F(w)\rangle\big ]  \big[F(w') - \left< F(w') \right>\big] \right>  \nn\ .
\end{eqnarray}
Here $w$ and $w'$ are two distinct  values of $w$ in Fig.~\ref{f:example}. 
Since  $\Delta (w,w')$   only depends on the difference $w-w'$ we can improve the disorder average (average over experiments) by an additional translational average. 

Let us start with an exact solution for a toy model \cite{LeDoussalWiese2008a}, namely   a particle dragged  through a disordered force landscape. To apply the results of Ref.~\cite{LeDoussalWiese2008a} one needs to specify  the distribution of forces $F$. Since the {\em microscopic} forces   can be thought of as sums of random variables (neighboring monomers act together to generate these random forces),  and  assuming the central-limit theorem   applies,  forces are   Gauss-distributed, which in the terminology of \cite{LeDoussalWiese2008a} leads to the Gumbell universality class of extreme-value statistics, with  correlator  
\begin{eqnarray}
\label{Gumbell-resc}
\Delta(w)&=& m^4 \rho_{m}^2 \Delta_{\rm Gumbell}(w/\rho_m) \ ,\\
\label{Gumbell}
\Delta_{\rm Gumbell}(x) &:=& \frac{x^2}2 + \mbox{Li}_2\Big(1-\rme^{|x|}\Big) + \frac{\pi^2}6\ , \\
\label{rho}
\rho_{m} &=& \frac{1}{  m^2
   \sqrt{2\log
    ( m^{-2} )}
   }\ .
\label{Gumbell-resc3}
\end{eqnarray}
As shown below in section   \ref{s:Data-Analysis}  and on  Fig.~\ref{f:our-choice}, experimental data and theory agree well. 

While it is gratifying to have a theoretical prediction verified by an experiment, 
the implications here are much deeper, as the function $\Delta(w)$ is the central object of the field theory of disordered elastic manifolds of inner dimension $d$ (here $d=0$), e.g.\ a magnetic domain wall in a bulk magnet ($d=2$), or a contact line ($d=1$). These  systems are governed by an equation of motion for the domain wall or line $u$, parameterized by an internal  coordinate $x$ and time $t$, 
\be\label{EOM}
\partial_t u(x,t) = \nabla^2 u(x,t) + m^2 \left[w -u(x,t) \right] + F\big (x,u(x,t)\big)\ . 
\end{equation}
 The force $F(w)$ is   the force acting on the center of mass.
Then  the {\em renormalized force-force correlator} \cite{LeDoussal2006b,LeDoussalWiese2006a,WieseLeDoussal2006}
\begin{equation}\label{Delta-def}
\Delta(w-w') := \frac{1}{L^d}\left< F (w) F(w') \right>^{\rm c}\ ,
\end{equation}
where $L$ is the linear size of the system, and $L^d$ its   volume, 
can be obtained from field theory\cite{NarayanDSFisher1993a,NattermannStepanowTangLeschhorn1992,LeDoussalWieseChauve2002,WieseLeDoussal2006}, based on the equation of motion \eq{EOM}. Field theory is a central tool in theoretical physics \cite{Zinn}, with applications ranging from elementary particle physics \cite{WeinbergBook123} to the fluctuations   observed around the critical point in liquid-gas transitions \cite{PhaseTransitionsAndCriticalPhenomenaSeries}. In all these cases, a set of {\em flow equations}  for a finite number of {\em coupling constants} is derived. 
These methods fail for disordered systems as those given by Eq.~\eq{EOM}. A way out was found by realizing that the flow 
for the coupling constants has to be generalized to   flow    for a function. This is known as  the {\em functional renormalization group} (FRG). The flow-equations take the form 
\begin{equation}\label{FRG-flow}
\partial_\ell \Delta(w) = -\frac{\rmd^2 }{\rmd w^2} \half\left[ \Delta(w)-\Delta(0)\right] ^2 + ...
\end{equation}
where the omitted terms are higher-order corrections (technically higher-loop terms \cite{Zinn,ChauveLeDoussalWiese2000a,LeDoussalWieseChauve2002,LeDoussalWieseChauve2003}), equivalent to an expansion in $\epsilon =4-d$ (with $d$ the dimension of the object). 
What came as a surprise was the realization that $\Delta(w)$ appearing in \Eq{FRG-flow}, when integrated from a microscopic scale to the length scale $\ell\equiv 1/m$ is the  disorder-force correlator measured via  \Eq{Delta-def} \cite{LeDoussal2006b,LeDoussalWiese2006a,WieseLeDoussal2006}.  
Measuring $\Delta (w)$ is thus   a key test \cite{MiddletonLeDoussalWiese2006,RossoLeDoussalWiese2006a,LeDoussalWieseMoulinetRolley2009} for the field theory of disordered systems. 
\begin{figure}[t]\setlength{\unitlength}{0.1\columnwidth}
\fboxsep0mm
{\begin{picture}(10,6.5)
\put(0,0){\fig{1}{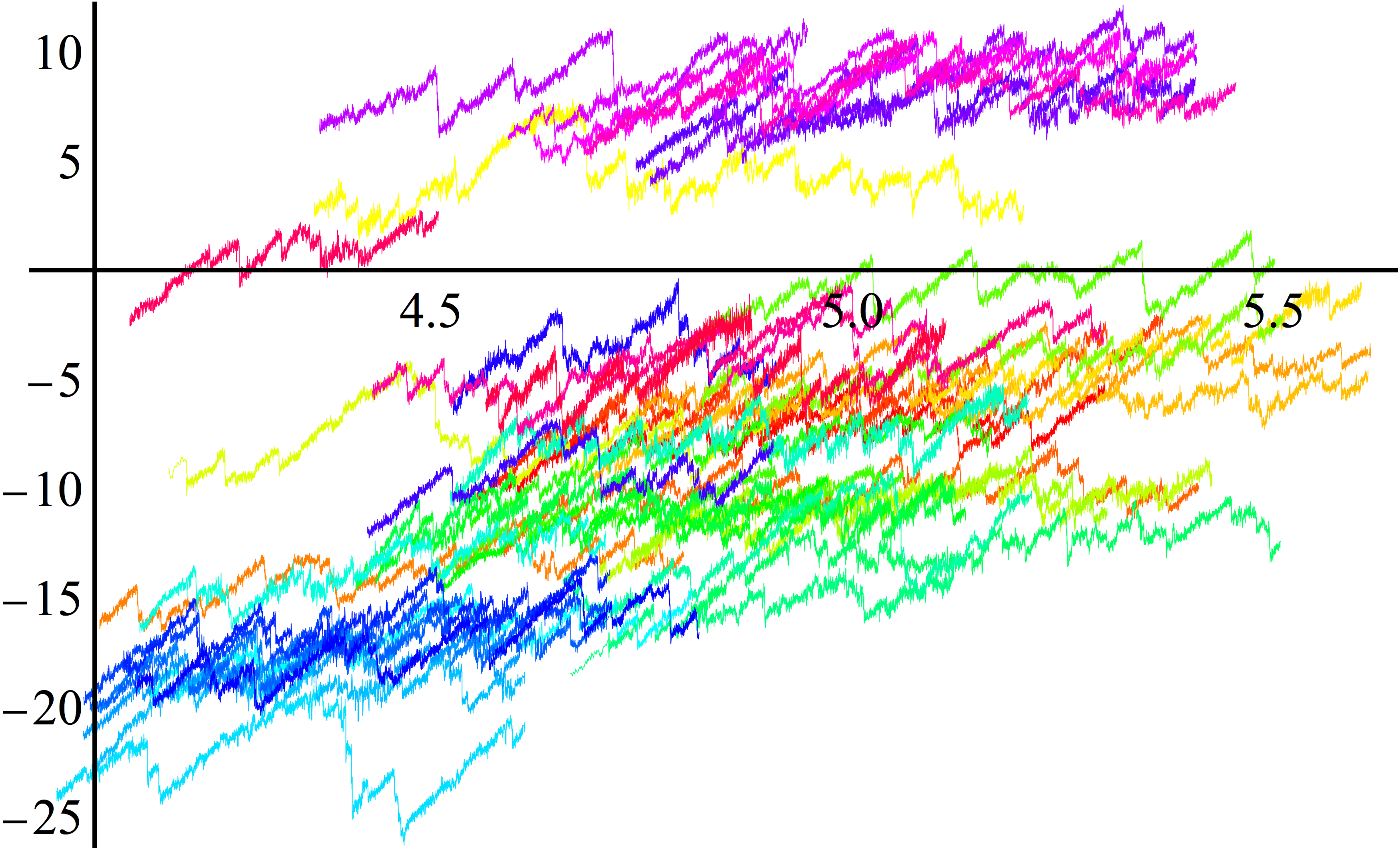}}
\put(9,4.3){$w[\mu\rm m]$}
\put(0,6.2){$F[\rm pN]$}
\end{picture}}
\caption{Force-extension curves restricted to the plateau region for one of our batches with 47 data sets. Curves are randomly displaced for better visualization. 
}
\label{f:aller-plateau4paper-Fw}
\end{figure}
The solution to Eq.~(\ref{FRG-flow})  (leading order in the expansion parameter $\epsilon$) reads 
\begin{eqnarray}\label{Delta-FTm1}
\Delta(w) &=& {\cal A} 
\, \Delta_{\rm FT}(w/\rho )\ , \\
\Delta_{\rm FT}(x) &=&-   
W\!\left(-\exp\!\big(\textstyle{-}\frac{x^2}{2}{-}1\big)\right)\ , 
\label{Delta-FT}
\end{eqnarray}
where $\cal A$ and $\rho$ are non-universal constants, and  the product-log 
$W(z)$ is the principal solution for $w$ in $z = w \rme^w$. 
Field theory also applies to the  experiment described above, which has (internal) dimension $d=0$ (the single degree of freedom   is the number of the last unpeeled monomer). 
As $\epsilon=4$ is large, we expect Eqs.~\eq{Delta-FTm1}-\eq{Delta-FT} to give a decent approximation for the experiment, but also to show differences. This is confirmed below, see section \ref{s:Data-Analysis} and Fig.~\ref{f:our-choice}.

\begin{figure}[t]\setlength{\unitlength}{0.1\columnwidth}
\fboxsep0mm
{\begin{picture}(10,6.88)
\put(0,0){\fig{1}{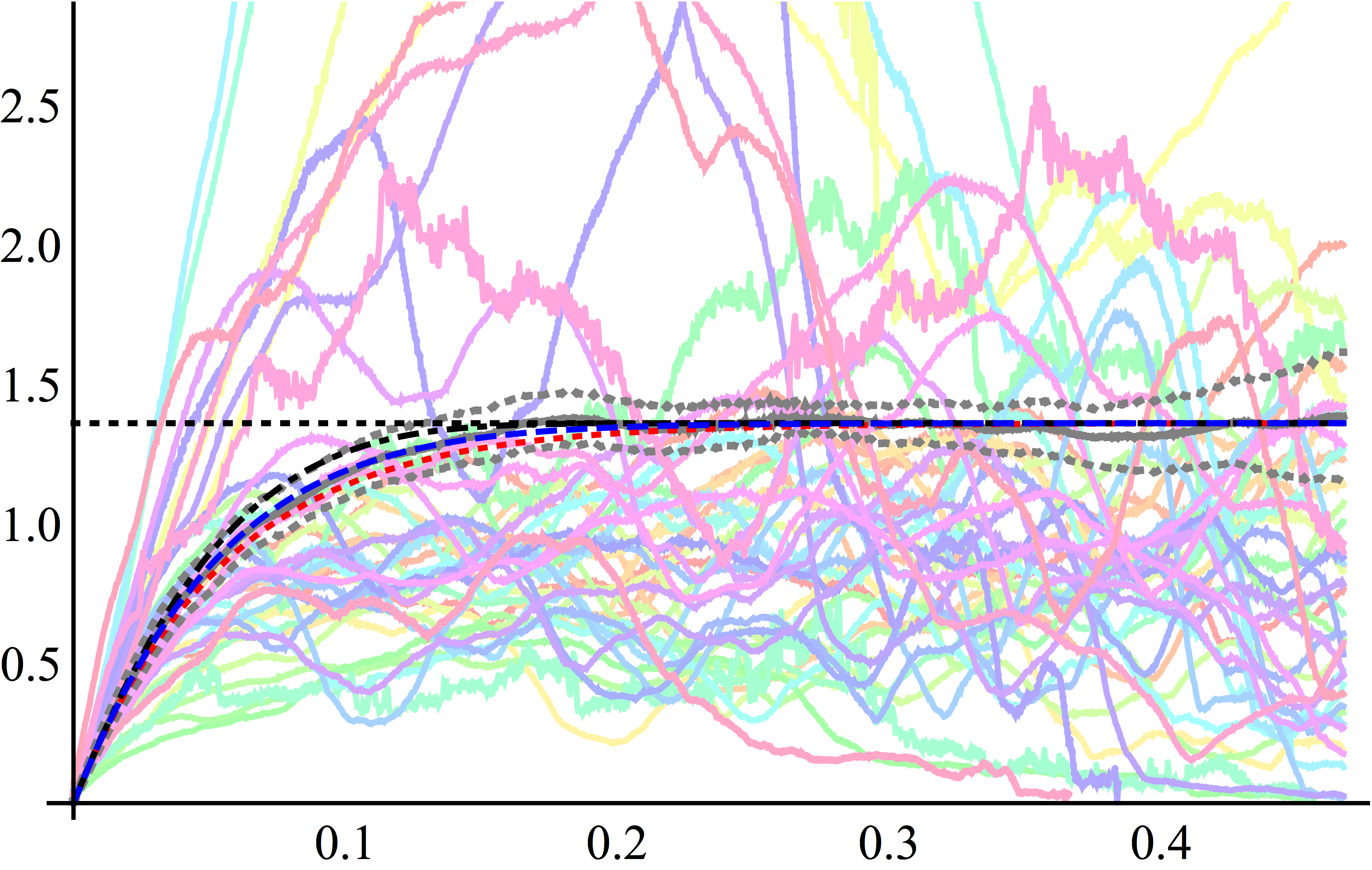}}
\put(9,0.1){$w[\mu\rm m]$}
\put(0,6.6){$\Delta(0)-\Delta(w)~[\rm pN^{2}]$}
\end{picture}}
\caption{Estimation of $ \Delta(0)-\Delta(w)$ from one of our batches with 47 datasets, compared to three theoretical curves: pure exponential decay (red), 1-loop FRG (black dot-dashed), and toy model (blue dashed).
}
\label{f:Delta-intermediate}
\end{figure}

\section{Data-Analysis and   Universal Signal}
\label{s:Data-Analysis}

We measure the force-extension curve in an RNA-DNA-unzipping experiment \cite{BercyThesis,MelkonyanThesis,MelkonyanBercyBizebardBockelmann2019}, retaining from the force-extension curve shown on Fig.~\ref{f:example} only the plateau part (in red). This experiment was repeated 163 times. From one of the batches with 47 data sets, we show the retained plateaux  on Fig.~\ref{f:aller-plateau4paper-Fw}. In order to minimize statistical errors, we measure the combination 
$\Delta(0) - \Delta(w) = \frac{1}2 \left< [ F(u+w)-F(u) ]^{2} \right>^{\!\rm c}  $. 
This average is more stable experimentally, since there seems   to be a small drift in the data (visible on Fig.~\ref{f:aller-plateau4paper-Fw}); the latter may be induced by a slightly diminishing effective stiffness $m^{2}$ upon opening the strands, even though this   is invisible on Fig.~\ref{f:example}. 

On figure \ref{f:Delta-intermediate}, we show the combination $\Delta(0)-\Delta(w)$ as defined by \Eq{Delta-def}, for each of the force-extension curves of Fig.~\ref{f:aller-plateau4paper-Fw}, with the shaded colors identical to those of Fig.~\ref{f:aller-plateau4paper-Fw}. Strong statistical fluctuations are visible. 
Their mean, in solid grey, is compared to three theoretical curves: The leading-order field theory result \eq{Delta-FT} (black dot-dashed line), the Gumbell result \eq{Gumbell-resc} (blue, dashed), and an exponentially decaying function (red, dotted). 
There are two unknown scales, equivalent to a rescaling of $w$ and $\Delta$. Since the slope at the origin can be measured precisely, we rescale all theoretical functions to have the same slope \cite{note-noisy-data}. The remaining parameter is the behavior of $\Delta(0)-\Delta(w)$ for large $w$, which is adjusted visually. In dotted gray we show our estimates of the absolute error bars, obtained by     resampling, as explained in App.~\ref{s:Data-analysis+error-bars}.  

The result (of these partial data)   favors the theoretical prediction \eq{Gumbell-resc}, while the estimated error bars are seemingly rather large. The reason for the latter is that the main statistical fluctuations come from the amplitude multiplying $\Delta(w)$.  Measuring the   statistical error of $\Delta(w)/\Delta(0)$ estimates the error of the shape only.  It is further reduced by  using all our data, and is indicated by the green shaded region  on our final curve on Fig.~\ref{f:our-choice}.
The agreement of the theory and the experimental data is excellent,   better than expected from the single measurements of Fig.~\ref{f:Delta-intermediate}. 
This strongly indicates that the {\em universal physics} behind the depinning transition is  robust.

\section{Scales and Interpretation}
\begin{figure}[t]\setlength{\unitlength}{0.1\columnwidth}
\fboxsep0mm
{\begin{picture}(10,8.7)
\put(3,2.5){\fig{0.7}{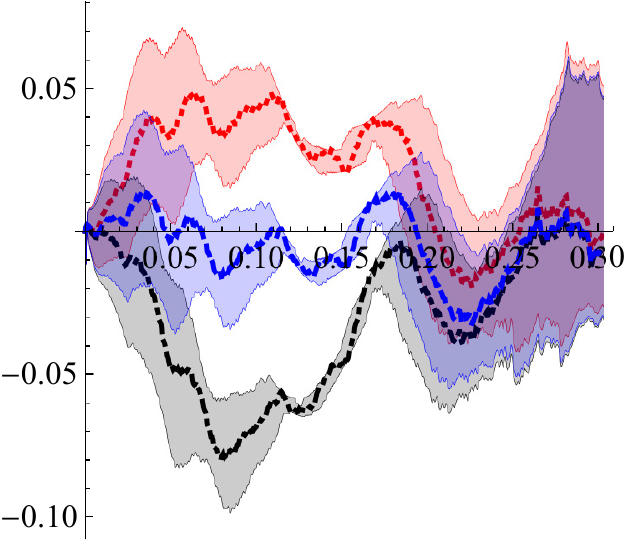}}
\put(0,0){\Fig{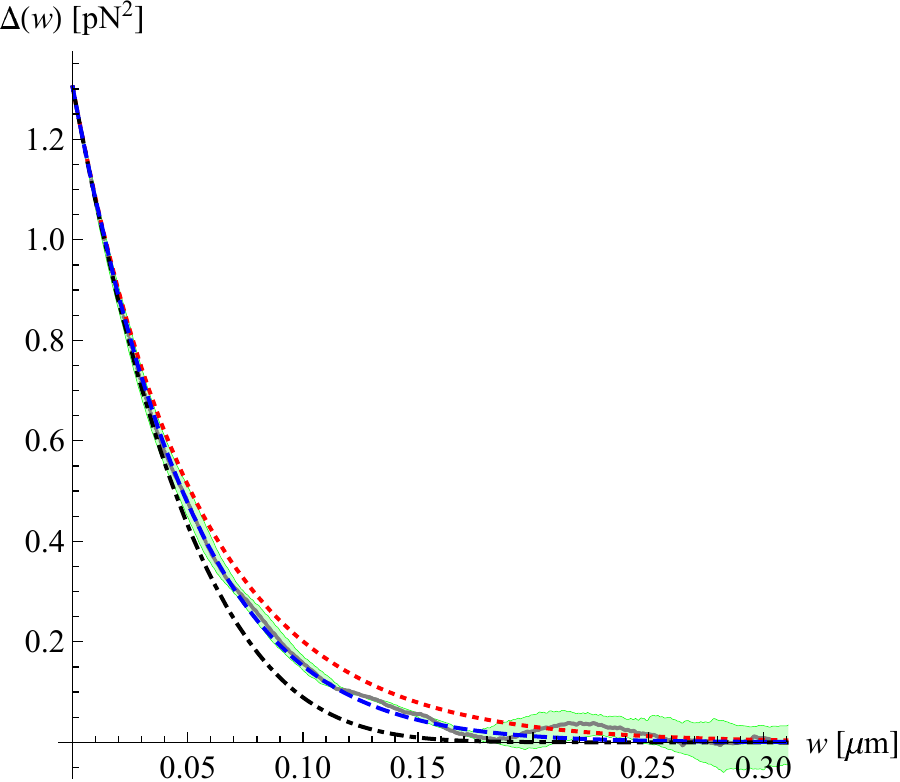}}
\end{picture}}
\caption{Measurements of $\Delta(w)$ (in grey), with $1-\sigma$ error bars (green shaded), compared to three theoretical curves: pure exponential decay (dotted red), 1-loop FRG (black dot-dashed), and toy model (blue dashed). Inset:  theoretical curves with the data subtracted (same color code). The blue curve is   the closest to the data. 
}
\label{f:our-choice}
\end{figure}

Our final result for $\Delta(w)$, given by the grey solid line  on Fig.~\ref{f:our-choice}, is in remarkable agreement with the analytical result \eq{Gumbell-resc}. What does this mean?
Consider   Fig.~\ref{f:example}, where the force grows linearly, interrupted by sudden   drops of size $\delta F$. One can show \cite{LeDoussalWiese2008c} that the derivative of the function $\Delta(w)$ at the origin is related to a moment ratio of force drops \cite{note-F-w}
\begin{equation}
|\Delta'(0^{+})| = m^{2} \delta F_{m}\ , \qquad  \delta F_{m} =  \frac{\left< \delta F^{2} \right>}{2\left< \delta F\right>}\ .
\end{equation}
Our experiments yield 
$m^{2}= 55\pm 5 \, \rm pN/\mu m$ (see Fig.~\ref{1})
leading to 
 $\delta F_{m} = 0.43 \pm 0.05 \,\rm pN$, and to a correlation length  $\xi = 0.055\pm0.005 \mu \rm m \simeq 186$ base pairs. 
This is roughly consistent with the 9 force drops identifiable on figure \ref{f:example}.
The driving velocity was varied from 5 to 7 $\rm n m/s$, where no statistically significant difference was observed for $\Delta(w)$.

%
These measurements indicate a serious challenge for peeling experiments using optical tweezers: As force-force correlations decay on a scale of about 200 bases, which is about $1/15$ of the length of the ribosomal RNA, events can be resolved with that resolution. As \Eq{rho} shows, this resolution increases with the  stiffness $m^2$. The key to    a high resolution  is thus a stiff construction and a well-aligned trap: If the trap is not optimally aligned,   showing the critical force at a say $20 \%$ smaller value, the resolution suffers according to Eqs.~\eq{Gumbell-resc}-\eq{Gumbell-resc3} by an even larger amount. 
Another possibility to increase the stiffness is to use shorter constructions.

Our   system maximises    force differences, and thus the measured signal $\Delta(w)$, as the  two possible parings CG 
and AT/AU   with different binding energies  appear almost in the same proportion \cite{BercyThesis}.
The binding enthalpy $\delta \cal H$ per hydrogen bond is about 
$2\, \rm kcal/mol$  \cite{BreslauerFrankBlockerMarky1986}, which gives $5\, \rm kcal/mol$ on average for the two possible pairings. 
This is augmented by stacking energies between the DNA and RNA strand to     $9.1\,\rm kcal/mol$ \cite{SugimotoNakanoKatohMatsumuraNakamutaOhmichiYoneyamaSasaki1995}. 
 The mean free energies $\delta \cal G$ are   1.5 kcal/mol \cite{SugimotoNakanoKatohMatsumuraNakamutaOhmichiYoneyamaSasaki1995}.
The difference in the number of hydrogen bonds  gives binding energy fluctuations of $1\, \rm kcal/mol$, which is $11\%$ of the total binding energy.  
The measured force $F_{\rm c}$ on the plateau is   $60\,\rm pN$, which leads to a microscopic force fluctuation of 
$\rmd F  =   6.6 \rm pN$. In the experiment, we observe d$F$ = 1.14 pN. Why is this value so low? 
If we use the correlation length of 186 base pairs, equate it with $\rho$ in Eqs.~\eq{Gumbell-resc}-\eq{rho}, this predicts $m^2=1.5\times 10^{-3}$, and 
$\log(m^{-2})=6.5$. This reduces $\rmd F$ from 6.6 pN to 2.6 pN, about 2 times larger than in the experiment.  
Several factors may contribute to  a further reduction:  thermal fluctuations, and
 the appearance of base {\em pairs} in the   free-energy estimates   \cite{SugimotoNakanoKatohMatsumuraNakamutaOhmichiYoneyamaSasaki1995}.
Finally, we did not discard ``bad'' samples, i.e.\ samples with vanishing signal, since any filtering risks to introduce a systematic bias. While this does not change the shape of $\Delta(w)$, it   underestimates its amplitude. 
Thus our measurements agree qualitatively  with this crude estimate, and more should not be expected.

The function $\Delta(w)$ was also measured in contact-line depinning \cite{LeDoussalWieseMoulinetRolley2009}, and studied numerically for magnetic domain walls \cite{MiddletonLeDoussalWiese2006,RossoLeDoussalWiese2006a}. For all these systems   thermal fluctuations play no role.  
In our experiment,  {\em thermal fluctuations} are clearly visible, both in an additional white noise for each data point, as in  a reduction of  
$\delta {\cal H}\approx 9.1\,\rm kcal/mol$     
to $\delta {\cal G}\approx 1.5 \,\rm kcal/mol$. Expectations  in the literature are that thermal fluctuations may round the cusp \cite{WieseLeDoussal2006,ChauveGiamarchiLeDoussal2000}, a feature not visible here. 
How can this be reconciled? There are several distinct entropic contributions:  Conformational  entropy of unbound strands, a higher entropy for the salt ions for unbound strands, and binding entropy, as a bond can  either  be  open or closed. The first two ones are large and reduce $\delta \cal H$ quite substantially to $\delta \cal G$; the last one is small: as $k_{{\rm B}}T\approx 0.6 \rm kcal/mol$, the probability that a bond is broken by  thermal fluctuations is $p_{  T}=\rme^{{-\delta {\cal G}/k_{\rm B}T}}$, with values ranging from  $8 \times 10^{-3}$ to $0.7$ using the binding free energies $\delta \cal G$ of \cite{SugimotoNakanoKatohMatsumuraNakamutaOhmichiYoneyamaSasaki1995}. Thus at most a few bonds can be opened  by thermal fluctuations, and using Eqs.~\eq{Gumbell-resc} and \eq{Delta-FT} based on zero-temperature depinning is justified.

One should be able to extract $\Delta(w)$ also from the unzipping \cite{unzipping} of a hairpin. Interestingly, experiments report that the  scaling of  \Eq{Gumbell-resc3} is   replaced by $\rho_m \sim m^{-4/3}$ \cite{HuguetFornsRitort2009}, a clear signature of a different universality class, namely   ``random-field'' disorder in equilibrium \cite{last-note}. This scenario is possible through the much stronger effective stiffness $m^2$ there. Equilibrium is observed experimentally through a vanishing hysteresis curve.

While the sequence used in the experiments is extracted from ribosomal RNA, thus is {\em not random}, the measured function $\Delta(w)$ agrees to a good precision  with that obtained for a {\em random} sequence. In conclusion, universal physics can emerge even in a specific biological system, and for a relatively short sequence.

%
%

\appendix

\section{Data analysis and error-estimates}
\label{s:Data-analysis+error-bars}

Protocal and error-estimates: 
Define for a data-set ${\cal D}_i$, with $i=1,...,n$ and $n$ the total number of   force-extension curves,  the set-average
\begin{eqnarray}
 {N_i(w)} &:=& \sum_{u\in {\cal D}_i}\\
{Q}_i(w) &:=& \frac{1}{N_i(w)} \sum_{u\in {\cal D}_i}  \Big[ F(u+w)-F(u)\Big]^{2}   \\
{M}_i(w) &:=& \frac{1}{N_i(w)} \sum_{u\in {\cal D}_i}  \Big[ F(u+w)-F(u)\Big]   \\
{Q}_i^{\rm c}(w) &:=& {Q}_i(w) -{M}_i(w)^2  
\end{eqnarray}
The above sums run over all values $u$, for which exists a pair $F(u+w)$ and $F(u)$; ${N}_i(w)$ is the number of such pairs. 
Our best estimate for the force-force correlator then is
\be
\left< \Big[ F(u+w)-F(u)\Big]^{2} \right>^{\!\!\rm c} =\frac{ \sum _i  Q_i^{\rm c} (w) N_{i}(w) }{\sum_i N_i(w)}\ .
\end{equation}
The fluctuations of the data shown on Fig.~\ref{f:Delta-intermediate} are very large, making error-estimates difficult. 
We used a statistical  resampling technique:
 Randomly divide all datasets ${\cal D}_{i}$ into two parts, ${\cal P}_1$ and  ${\cal P}_2$.
Define
\begin{eqnarray}
N_{{\cal P}_1}(w) &:=& \sum _{i\in {\cal P}_1} N_i(w)\ , \\
Q_{{\cal P}_1}^{\rm c} (w) &:=& \frac1{N_{{\cal P}_1}(w)} \sum _{i\in {\cal P}_1}   {Q_i^{\rm c} (w) N_{i}(w) } \ .
\end{eqnarray}
A similar definition holds for ${\cal P}_2$. 
Then for each $w$ measure the variance of the partial means $Q_{{\cal P}_1}^{\rm c} (w)$ and $Q_{{\cal P}_2}^{\rm c} (w)$. Finally,  average over all   partitions $\Pi_i, \{1,...,n\} \to {\cal P}_1,{\cal P}_2$. 
In practice, it is enough to take   $N_{\rm p}=100$ random partitions. The error estimate then is
\begin{eqnarray}
\sigma^2(w) &:=&\frac1{N_{\rm p}} \sum_{\Pi_i} \left< \frac12\sum_{k=1}^2 \left[ Q_{{\cal P}_k(\Pi_i)}^{\rm c} (w)  - Q^{\rm c} (w)\right]^2 \right>, ~~~~~~~ \\
N_{\rm p} &:=&\sum_{\Pi_i}.
\end{eqnarray}
We can also define the set of all $2 N_{\rm p} $ partial means,
\be\label{A10}
{\cal A} (w) := \bigcup_{\Pi_i} \bigcup_{k=1,2}   Q_{{\cal P}_k(\Pi_i)}^{\rm c}(w) \ .
\end{equation}
We find that our analysis is consistent, with 
\be
\mbox{var} \Big({\cal A} (w) \Big) \approx     \sigma^2(w) \ .
\end{equation}
These error estimates  are absolute errors, presented on Fig.~\ref{f:Delta-intermediate}. 
To obtain the error estimate given on Fig.~\ref{f:our-choice},  the partial means \eq{A10} where rescaled such that their $w$-integrals  equal the $w$-integral over all samples. This takes out amplitude fluctuations, reducing the errors to errors of the shape.

\medskip

\section{Check on test data}
We generated test data according to the following protocol:  
For each real data set, sample an Ornstein-Uhlenbeck process of the same length, with 
   mean 
$F_m =F_c$, variance  $\Delta(0)$,   and correlation length   $\xi$ as measured. This is achieved by the stochastic process
\begin{eqnarray}
&& F(w+\delta w) = F(w) + \zeta(w) \sqrt{\frac{\Delta(0)}{\xi}} + \frac{F_m-F(w)}{\xi} \ ,~~~~~\quad \\
&& \left< \zeta(w) \zeta(w')\right> = \delta_{w,w'} \ .  ~~~
\end{eqnarray}
This gives a first set of  test data. For a second set, we add an additional white noise in the $x$-direction, with $\delta x \in  \{ -1,0,1\}$, in units of the resolution of the measuring mashine. For a third set, we added a Gauss-distribution of mean zero and width 1 to the force signal. 
By construction, these test-data are exponentially correlated 
\be
\left< F(w) F(w')\right>^{\rm c} = \Delta(0)\rme^{-|w-w'|/\xi}\ , 
\end{equation}
with additional noise for sets 2 and 3. They should thus approach the red dotted curve of Fig.~\ref{f:our-choice}. This is indeed observed, with an appropriate  estimate  for the error bars.

\enlargethispage{1cm}

\ifx\doi\undefined
\providecommand{\doi}[2]{\href{http://dx.doi.org/#1}{#2}}
\else
\renewcommand{\doi}[2]{\href{http://dx.doi.org/#1}{#2}}
\fi
\providecommand{\link}[2]{\href{#1}{#2}}
\providecommand{\arxiv}[1]{\href{http://arxiv.org/abs/#1}{#1}}
\providecommand{\mrnumber}[1]{\href{https://mathscinet.ams.org/mathscinet/search/publdoc.html?pg1=MR&s1=#1&loc=fromreflist}{MR#1}}


\begin{thebibliography}{38}%
\makeatletter
\providecommand \@ifxundefined [1]{%
 \@ifx{#1\undefined}
}%
\providecommand \@ifnum [1]{%
 \ifnum #1\expandafter \@firstoftwo
 \else \expandafter \@secondoftwo
 \fi
}%
\providecommand \@ifx [1]{%
 \ifx #1\expandafter \@firstoftwo
 \else \expandafter \@secondoftwo
 \fi
}%
\providecommand \natexlab [1]{#1}%
\providecommand \enquote  [1]{``#1''}%
\providecommand \bibnamefont  [1]{#1}%
\providecommand \bibfnamefont [1]{#1}%
\providecommand \citenamefont [1]{#1}%
\providecommand \href@noop [0]{\@secondoftwo}%
\providecommand \href [0]{\begingroup \@sanitize@url \@href}%
\providecommand \@href[1]{\@@startlink{#1}\@@href}%
\providecommand \@@href[1]{\endgroup#1\@@endlink}%
\providecommand \@sanitize@url [0]{\catcode `\\12\catcode `\$12\catcode
  `\&12\catcode `\#12\catcode `\^12\catcode `\_12\catcode `\%12\relax}%
\providecommand \@@startlink[1]{}%
\providecommand \@@endlink[0]{}%
\providecommand \url  [0]{\begingroup\@sanitize@url \@url }%
\providecommand \@url [1]{\endgroup\@href {#1}{\urlprefix }}%
\providecommand \urlprefix  [0]{URL }%
\providecommand \Eprint [0]{\href }%
\providecommand \doibase [0]{https://doi.org/}%
\providecommand \selectlanguage [0]{\@gobble}%
\providecommand \bibinfo  [0]{\@secondoftwo}%
\providecommand \bibfield  [0]{\@secondoftwo}%
\providecommand \translation [1]{[#1]}%
\providecommand \BibitemOpen [0]{}%
\providecommand \bibitemStop [0]{}%
\providecommand \bibitemNoStop [0]{.\EOS\space}%
\providecommand \EOS [0]{\spacefactor3000\relax}%
\providecommand \BibitemShut  [1]{\csname bibitem#1\endcsname}%
\let\auto@bib@innerbib\@empty
\bibitem [{\citenamefont {Stephens}\ \emph {et~al.}(2015)\citenamefont
  {Stephens}, \citenamefont {Lee}, \citenamefont {Faghri}, \citenamefont
  {Campbell}, \citenamefont {Zhai}, \citenamefont {Efron}, \citenamefont
  {Iyer}, \citenamefont {Schatz}, \citenamefont {Sinha},\ and\ \citenamefont
  {Robinson}}]{StephensLeeFaghriCampbellZhaiEfronIyerSchatzSinhaRobinson2015}%
  \BibitemOpen
  \bibfield  {author} {\bibinfo {author} {\bibfnamefont {Z.}~\bibnamefont
  {Stephens}}, \bibinfo {author} {\bibfnamefont {S.}~\bibnamefont {Lee}},
  \bibinfo {author} {\bibfnamefont {F.}~\bibnamefont {Faghri}}, \bibinfo
  {author} {\bibfnamefont {R.}~\bibnamefont {Campbell}}, \bibinfo {author}
  {\bibfnamefont {C.}~\bibnamefont {Zhai}}, \bibinfo {author} {\bibfnamefont
  {M.}~\bibnamefont {Efron}}, \bibinfo {author} {\bibfnamefont
  {R.}~\bibnamefont {Iyer}}, \bibinfo {author} {\bibfnamefont {M.}~\bibnamefont
  {Schatz}}, \bibinfo {author} {\bibfnamefont {S.}~\bibnamefont {Sinha}},\ and\
  \bibinfo {author} {\bibfnamefont {G.}~\bibnamefont {Robinson}},\ }\bibfield
  {title} {\bibinfo {title} {Big data: Astronomical or genomical?},\ }\href
  {https://doi.org/10.1371/journal.pbio.1002195} {\bibfield  {journal}
  {\bibinfo  {journal} {PLOS Biology}\ }\textbf {\bibinfo {volume} {13}},\
  \bibinfo {pages} {1} (\bibinfo {year} {2015})}\BibitemShut {NoStop}%
%
\bibitem{unzipping}
In the literature, the word {\em peeling}  is used for the setup of Fig.~\ref{f:peeling},  where forces act along the helical axis from opposite  extremities of a duplex, and one of the two strands peels off.  {\em Unzipping} denotes an alternative setup where the right bead of Fig.~\ref{f:peeling}
 is attached to the free end of the upper strand. 
%
\bibitem [{\citenamefont {Ashkin}(1970)}]{Ashkin1970}%
  \BibitemOpen
  \bibfield  {author} {\bibinfo {author} {\bibfnamefont {A.}~\bibnamefont
  {Ashkin}},\ }\bibfield  {title} {\bibinfo {title} {Acceleration and trapping
  of particles by radiation pressure},\ }\href
  {https://doi.org/10.1103/PhysRevLett.24.156} {\bibfield  {journal} {\bibinfo
  {journal} {Phys. Rev. Lett.}\ }\textbf {\bibinfo {volume} {24}},\ \bibinfo
  {pages} {156} (\bibinfo {year} {1970})}\BibitemShut {NoStop}%
\bibitem [{\citenamefont {Bercy}(2015)}]{BercyThesis}%
  \BibitemOpen
  \bibfield  {author} {\bibinfo {author} {\bibfnamefont {M.}~\bibnamefont
  {Bercy}},\ } {\bibinfo {title}
  {\href{http://tel.archives-ouvertes.fr/tel-01409699}{Structures secondaires
  dans l'{ARN}: une \'etude par mesure de forces sur mol\'ecules uniques}}},\
  \href@noop {} {Ph.D. thesis},\ \bibinfo  {school} {PSL Research University}
  (\bibinfo {year} {2015})\BibitemShut {NoStop}%
\bibitem [{\citenamefont {Melkonyan}(2018)}]{MelkonyanThesis}%
  \BibitemOpen
  \bibfield  {author} {\bibinfo {author} {\bibfnamefont {L.}~\bibnamefont
  {Melkonyan}},\ }\emph {\bibinfo {title}
  {\href{http://www.theses.fr/s176114}{Early stages of ribosome assembly,
  studied by single-molecule force measurements}}},\ \href@noop {} {Ph.D.
  thesis},\ \bibinfo  {school} {PSL Research University} (\bibinfo {year}
  {2018})\BibitemShut {NoStop}%
%
\bibitem [{\citenamefont {Gotta}\ \emph {et~al.}(1991)\citenamefont {Gotta},
  \citenamefont {Miller},\ and\ \citenamefont
  {French}}]{GottaMillerFrench1991}%
  \BibitemOpen
  \bibfield  {author} {\bibinfo {author} {\bibfnamefont {S.}~\bibnamefont
  {Gotta}}, \bibinfo {author} {\bibfnamefont {O.}~\bibnamefont {Miller}},\ and\
  \bibinfo {author} {\bibfnamefont {S.}~\bibnamefont {French}},\ }\bibfield
  {title} {\bibinfo {title} {{rRNA} transcription rate in escherichia coli},\
  }\href {https://doi.org/10.1128/jb.173.20.6647-6649.1991} {\bibfield
  {journal} {\bibinfo  {journal} {J. Bacteriol.}\ }\textbf {\bibinfo {volume}
  {173}},\ \bibinfo {pages} {6647} (\bibinfo {year} {1991})}\BibitemShut
  {NoStop}%
%
\bibitem [{\citenamefont {Melkonyan}\ \emph {et~al.}(2019)\citenamefont
  {Melkonyan}, \citenamefont {Bercy}, \citenamefont {Bizebard},\ and\
  \citenamefont {Bockelmann}}]{MelkonyanBercyBizebardBockelmann2019}%
  \BibitemOpen
  \bibfield  {author} {\bibinfo {author} {\bibfnamefont {L.}~\bibnamefont
  {Melkonyan}}, \bibinfo {author} {\bibfnamefont {M.}~\bibnamefont {Bercy}},
  \bibinfo {author} {\bibfnamefont {T.}~\bibnamefont {Bizebard}},\ and\
  \bibinfo {author} {\bibfnamefont {U.}~\bibnamefont {Bockelmann}},\ }\bibfield
   {title} {\bibinfo {title} {Overstretching double-stranded {RNA},
  double-stranded {DNA}, and {RNA-DNA} duplexes},\ }\href
  {https://doi.org/10.1016/j.bpj.2019.07.003} {\bibfield  {journal} {\bibinfo
  {journal} {Biophys. J.}\ }\textbf {\bibinfo {volume} {117}},\ \bibinfo
  {pages} {509} (\bibinfo {year} {2019})}\BibitemShut {NoStop}%
\bibitem [{\citenamefont {L\'eger}\ \emph {et~al.}(1999)\citenamefont
  {L\'eger}, \citenamefont {Romano}, \citenamefont {Sarkar}, \citenamefont
  {Robert}, \citenamefont {Bourdieu}, \citenamefont {Chatenay},\ and\
  \citenamefont {Marko}}]{LegerRomanoSarkarRobertBourdieuChatenayMarko1999}%
  \BibitemOpen
  \bibfield  {author} {\bibinfo {author} {\bibfnamefont {J.~F.}\ \bibnamefont
  {L\'eger}}, \bibinfo {author} {\bibfnamefont {G.}~\bibnamefont {Romano}},
  \bibinfo {author} {\bibfnamefont {A.}~\bibnamefont {Sarkar}}, \bibinfo
  {author} {\bibfnamefont {J.}~\bibnamefont {Robert}}, \bibinfo {author}
  {\bibfnamefont {L.}~\bibnamefont {Bourdieu}}, \bibinfo {author}
  {\bibfnamefont {D.}~\bibnamefont {Chatenay}},\ and\ \bibinfo {author}
  {\bibfnamefont {J.~F.}\ \bibnamefont {Marko}},\ }\bibfield  {title} {\bibinfo
  {title} {Structural transitions of a twisted and stretched dna molecule},\
  }\href {https://doi.org/10.1103/PhysRevLett.83.1066} {\bibfield  {journal}
  {\bibinfo  {journal} {Phys. Rev. Lett.}\ }\textbf {\bibinfo {volume} {83}},\
  \bibinfo {pages} {1066} (\bibinfo {year} {1999})}\BibitemShut {NoStop}%
\bibitem [{\citenamefont {Strick}\ \emph {et~al.}(1996)\citenamefont {Strick},
  \citenamefont {Allemand}, \citenamefont {Bensimon}, \citenamefont
  {Bensimon},\ and\ \citenamefont
  {Croquette}}]{StrickAllemandBensimonBensimonCroquette1996}%
  \BibitemOpen
  \bibfield  {author} {\bibinfo {author} {\bibfnamefont {T.}~\bibnamefont
  {Strick}}, \bibinfo {author} {\bibfnamefont {J.-F.}\ \bibnamefont
  {Allemand}}, \bibinfo {author} {\bibfnamefont {D.}~\bibnamefont {Bensimon}},
  \bibinfo {author} {\bibfnamefont {A.}~\bibnamefont {Bensimon}},\ and\
  \bibinfo {author} {\bibfnamefont {V.}~\bibnamefont {Croquette}},\ }\bibfield
  {title} {\bibinfo {title} {The elasticity of a single supercoiled dna
  molecule},\ }\href {https://doi.org/10.1126/science.271.5257.1835} {\bibfield
   {journal} {\bibinfo  {journal} {Science}\ }\textbf {\bibinfo {volume}
  {271}},\ \bibinfo {pages} {1835} (\bibinfo {year} {1996})}\BibitemShut
  {NoStop}%
\bibitem [{\citenamefont {De~Vlaminck}\ and\ \citenamefont
  {Dekker}(2012)}]{DeVlaminckDekker2012}%
  \BibitemOpen
  \bibfield  {author} {\bibinfo {author} {\bibfnamefont {I.}~\bibnamefont
  {De~Vlaminck}}\ and\ \bibinfo {author} {\bibfnamefont {C.}~\bibnamefont
  {Dekker}},\ }\bibfield  {title} {\bibinfo {title} {Recent advances in
  magnetic tweezers},\ }\href
  {https://doi.org/10.1146/annurev-biophys-122311-100544} {\bibfield  {journal}
  {\bibinfo  {journal} {Ann. Rev. Biophys.}\ }\textbf {\bibinfo {volume}
  {41}},\ \bibinfo {pages} {453} (\bibinfo {year} {2012})},\ \bibinfo {note}
  {pMID: 22443989}\BibitemShut {NoStop}%
\bibitem [{\citenamefont {Nelson}(2005)}]{Nelson2005}%
  \BibitemOpen
  \bibfield  {author} {\bibinfo {author} {\bibfnamefont {D.}~\bibnamefont
  {Nelson}},\ }\bibfield  {title} {\bibinfo {title} {Statistical physics of
  unzipping {DNA}},\ }in\ \href@noop {} {\emph {\bibinfo {booktitle} {Forces,
  Growth and Form in Soft Condensed Matter: At the Interface between Physics
  and Biology}}},\ \bibinfo {editor} {edited by\ \bibinfo {editor}
  {\bibfnamefont {A.~T.}\ \bibnamefont {Skjeltorp}}\ and\ \bibinfo {editor}
  {\bibfnamefont {A.~V.}\ \bibnamefont {Belushkin}}}\ (\bibinfo  {publisher}
  {\href{https://link.springer.com/chapter/10.1007/1-4020-2340-5_4}{Springer
  Netherlands}},\ \bibinfo {address} {Dordrecht},\ \bibinfo {year} {2005})\
  pp.\ \bibinfo {pages} {65--92}\BibitemShut {NoStop}%
\bibitem [{\citenamefont {Barkhausen}(1919)}]{Barkhausen1919}%
  \BibitemOpen
  \bibfield  {author} {\bibinfo {author} {\bibfnamefont {H.}~\bibnamefont
  {Barkhausen}},\ }\bibfield  {title} {\bibinfo {title} {{Zwei mit Hilfe der
  neuen Verst\"arker entdeckte Erscheinungen}},\ }\href@noop {} {\bibfield
  {journal} {\bibinfo  {journal} {Phys. Z.}\ }\textbf {\bibinfo {volume}
  {20}},\ \bibinfo {pages} {401} (\bibinfo {year} {1919})}\BibitemShut
  {NoStop}%
%
\bibitem{SethnaDahmenMyers2001}
J.P. Sethna, K.A. Dahmen  and C.R. Myers,
\newblock {Crackling noise},
\newblock \doi{10.1038/35065675}{\rm Nature {\bf 410}, 242--250 (2001).}
%
\bibitem [{\citenamefont {Doussal}\ \emph {et~al.}(2009)\citenamefont
  {Doussal}, \citenamefont {Wiese}, \citenamefont {Moulinet},\ and\
  \citenamefont {Rolley}}]{LeDoussalWieseMoulinetRolley2009}%
  \BibitemOpen
  \bibfield  {author} {\bibinfo {author} {\bibfnamefont {P.~L.}\ \bibnamefont
  {Doussal}}, \bibinfo {author} {\bibfnamefont {K.}~\bibnamefont {Wiese}},
  \bibinfo {author} {\bibfnamefont {S.}~\bibnamefont {Moulinet}},\ and\
  \bibinfo {author} {\bibfnamefont {E.}~\bibnamefont {Rolley}},\ }\bibfield
  {title} {\bibinfo {title} {Height fluctuations of a contact line: {A} direct
  measurement of the renormalized disorder correlator},\ }\href
  {https://doi.org/10.1209/0295-5075/87/56001} {\bibfield  {journal} {\bibinfo
  {journal} {EPL}\ }\textbf {\bibinfo {volume} {87}},\ \bibinfo {pages} {56001}
  (\bibinfo {year} {2009})},\ \Eprint {https://arxiv.org/abs/arXiv:0904.4156}
  {arXiv:0904.4156} \BibitemShut {NoStop}%
\bibitem [{\citenamefont {Gutenberg}\ and\ \citenamefont
  {Richter}(1956)}]{GutenbergRichter1956}%
  \BibitemOpen
  \bibfield  {author} {\bibinfo {author} {\bibfnamefont {B.}~\bibnamefont
  {Gutenberg}}\ and\ \bibinfo {author} {\bibfnamefont {C.}~\bibnamefont
  {Richter}},\ }\bibfield  {title} {\bibinfo {title} {{Earthquake magnitude,
  intensity, energy, and acceleration}},\ }\href
  {https://pubs.geoscienceworld.org/ssa/bssa/article-abstract/46/2/105/115777}
  {\bibfield  {journal} {\bibinfo  {journal} {Bulletin of the Seismological
  Society of America}\ }\textbf {\bibinfo {volume} {46}},\ \bibinfo {pages}
  {105} (\bibinfo {year} {1956})}\BibitemShut {NoStop}%
\bibitem [{\citenamefont {Blatter}\ \emph {et~al.}(1994)\citenamefont
  {Blatter}, \citenamefont {{Feigel'man}}, \citenamefont {Geshkenbein},
  \citenamefont {Larkin},\ and\ \citenamefont
  {Vinokur}}]{BlatterFeigelmanGeshkenbeinLarkinVinokur1994}%
  \BibitemOpen
  \bibfield  {author} {\bibinfo {author} {\bibfnamefont {G.}~\bibnamefont
  {Blatter}}, \bibinfo {author} {\bibfnamefont {M.}~\bibnamefont
  {{Feigel'man}}}, \bibinfo {author} {\bibfnamefont {V.}~\bibnamefont
  {Geshkenbein}}, \bibinfo {author} {\bibfnamefont {A.}~\bibnamefont
  {Larkin}},\ and\ \bibinfo {author} {\bibfnamefont {V.}~\bibnamefont
  {Vinokur}},\ }\bibfield  {title} {\bibinfo {title} {Vortices in
  high-temperature superconductors},\ }\href
  {https://doi.org/10.1103/RevModPhys.66.1125} {\bibfield  {journal} {\bibinfo
  {journal} {Rev. Mod. Phys.}\ }\textbf {\bibinfo {volume} {66}},\ \bibinfo
  {pages} {1125} (\bibinfo {year} {1994})}\BibitemShut {NoStop}%
\bibitem [{\citenamefont {Burridge}\ and\ \citenamefont
  {Knopoff}(1967)}]{BurridgeKnopoff1967}%
  \BibitemOpen
  \bibfield  {author} {\bibinfo {author} {\bibfnamefont {R.}~\bibnamefont
  {Burridge}}\ and\ \bibinfo {author} {\bibfnamefont {L.}~\bibnamefont
  {Knopoff}},\ }\bibfield  {title} {\bibinfo {title} {Model and theoretical
  seismicity},\ }\href@noop {} {\bibfield  {journal} {\bibinfo  {journal}
  {Bulletin of the Seismological Society of America}\ }\textbf {\bibinfo
  {volume} {57}},\ \bibinfo {pages} {341} (\bibinfo {year} {1967})}\BibitemShut
  {NoStop}%
\bibitem [{\citenamefont {Alessandro}\ \emph
  {et~al.}(1990{\natexlab{a}})\citenamefont {Alessandro}, \citenamefont
  {Beatrice}, \citenamefont {Bertotti},\ and\ \citenamefont
  {Montorsi}}]{AlessandroBeatriceBertottiMontorsi1990}%
  \BibitemOpen
  \bibfield  {author} {\bibinfo {author} {\bibfnamefont {B.}~\bibnamefont
  {Alessandro}}, \bibinfo {author} {\bibfnamefont {C.}~\bibnamefont
  {Beatrice}}, \bibinfo {author} {\bibfnamefont {G.}~\bibnamefont {Bertotti}},\
  and\ \bibinfo {author} {\bibfnamefont {A.}~\bibnamefont {Montorsi}},\
  }\bibfield  {title} {\bibinfo {title} {Domain-wall dynamics and {Barkhausen}
  effect in metallic ferromagnetic materials. {I. Theory}},\ }\href
  {https://doi.org/10.1063/1.346423} {\bibfield  {journal} {\bibinfo  {journal}
  {J. Appl. Phys.}\ }\textbf {\bibinfo {volume} {68}},\ \bibinfo {pages} {2901}
  (\bibinfo {year} {1990}{\natexlab{a}})}\BibitemShut {NoStop}%
\bibitem [{\citenamefont {Alessandro}\ \emph
  {et~al.}(1990{\natexlab{b}})\citenamefont {Alessandro}, \citenamefont
  {Beatrice}, \citenamefont {Bertotti},\ and\ \citenamefont
  {Montorsi}}]{AlessandroBeatriceBertottiMontorsi1990b}%
  \BibitemOpen
  \bibfield  {author} {\bibinfo {author} {\bibfnamefont {B.}~\bibnamefont
  {Alessandro}}, \bibinfo {author} {\bibfnamefont {C.}~\bibnamefont
  {Beatrice}}, \bibinfo {author} {\bibfnamefont {G.}~\bibnamefont {Bertotti}},\
  and\ \bibinfo {author} {\bibfnamefont {A.}~\bibnamefont {Montorsi}},\
  }\bibfield  {title} {\bibinfo {title} {{Domain-wall dynamics and Barkhausen
  effect in metallic ferromagnetic materials. II. Experiments}},\ }\href
  {https://doi.org/10.1063/1.346424} {\bibfield  {journal} {\bibinfo  {journal}
  {J. Appl. Phys.}\ }\textbf {\bibinfo {volume} {68}},\ \bibinfo {pages} {2908}
  (\bibinfo {year} {1990}{\natexlab{b}})}\BibitemShut {NoStop}%
\bibitem [{\citenamefont {Nattermann}\ \emph {et~al.}(1992)\citenamefont
  {Nattermann}, \citenamefont {Stepanow}, \citenamefont {Tang},\ and\
  \citenamefont {Leschhorn}}]{NattermannStepanowTangLeschhorn1992}%
  \BibitemOpen
  \bibfield  {author} {\bibinfo {author} {\bibfnamefont {T.}~\bibnamefont
  {Nattermann}}, \bibinfo {author} {\bibfnamefont {S.}~\bibnamefont
  {Stepanow}}, \bibinfo {author} {\bibfnamefont {L.-H.}\ \bibnamefont {Tang}},\
  and\ \bibinfo {author} {\bibfnamefont {H.}~\bibnamefont {Leschhorn}},\
  }\bibfield  {title} {\bibinfo {title} {Dynamics of interface depinning in a
  disordered medium},\ }\href {https://doi.org/10.1051/jp2:1992214} {\bibfield
  {journal} {\bibinfo  {journal} {J. Phys. II (France)}\ }\textbf {\bibinfo
  {volume} {2}},\ \bibinfo {pages} {1483} (\bibinfo {year} {1992})}\BibitemShut
  {NoStop}%
\bibitem [{\citenamefont {Narayan}\ and\ \citenamefont
  {Fisher}(1993)}]{NarayanDSFisher1993a}%
  \BibitemOpen
  \bibfield  {author} {\bibinfo {author} {\bibfnamefont {O.}~\bibnamefont
  {Narayan}}\ and\ \bibinfo {author} {\bibfnamefont {D.}~\bibnamefont
  {Fisher}},\ }\bibfield  {title} {\bibinfo {title} {Threshold critical
  dynamics of driven interfaces in random media},\ }\href
  {https://doi.org/10.1103/PhysRevB.48.7030} {\bibfield  {journal} {\bibinfo
  {journal} {Phys. Rev. B}\ }\textbf {\bibinfo {volume} {48}},\ \bibinfo
  {pages} {7030} (\bibinfo {year} {1993})}\BibitemShut {NoStop}%
\bibitem [{\citenamefont {Bucheli}\ \emph {et~al.}(1998)\citenamefont
  {Bucheli}, \citenamefont {Wagner}, \citenamefont {Geshkenbein}, \citenamefont
  {Larkin},\ and\ \citenamefont
  {Blatter}}]{BucheliWagnerGeshkenbeinLarkinBlatter1998}%
  \BibitemOpen
  \bibfield  {author} {\bibinfo {author} {\bibfnamefont {H.}~\bibnamefont
  {Bucheli}}, \bibinfo {author} {\bibfnamefont {O.}~\bibnamefont {Wagner}},
  \bibinfo {author} {\bibfnamefont {V.}~\bibnamefont {Geshkenbein}}, \bibinfo
  {author} {\bibfnamefont {A.}~\bibnamefont {Larkin}},\ and\ \bibinfo {author}
  {\bibfnamefont {G.}~\bibnamefont {Blatter}},\ }\bibfield  {title} {\bibinfo
  {title} {{$(4+N)$}-dimensional elastic manifolds in random media: a
  renormalization-group analysis},\ }\href@noop {} {\bibfield  {journal}
  {\bibinfo  {journal} {Phys. Rev. B}\ }\textbf {\bibinfo {volume} {57}},\
  \bibinfo {pages} {7642} (\bibinfo {year} {1998})}\BibitemShut {NoStop}%
\bibitem [{\citenamefont {Chauve}\ \emph {et~al.}(2001)\citenamefont {Chauve},
  \citenamefont {Doussal},\ and\ \citenamefont
  {Wiese}}]{ChauveLeDoussalWiese2000a}%
  \BibitemOpen
  \bibfield  {author} {\bibinfo {author} {\bibfnamefont {P.}~\bibnamefont
  {Chauve}}, \bibinfo {author} {\bibfnamefont {P.~L.}\ \bibnamefont
  {Doussal}},\ and\ \bibinfo {author} {\bibfnamefont {K.}~\bibnamefont
  {Wiese}},\ }\bibfield  {title} {\bibinfo {title} {Renormalization of pinned
  elastic systems: How does it work beyond one loop?},\ }\href
  {https://doi.org/10.1103/PhysRevLett.86.1785} {\bibfield  {journal} {\bibinfo
   {journal} {Phys. Rev. Lett.}\ }\textbf {\bibinfo {volume} {86}},\ \bibinfo
  {pages} {1785} (\bibinfo {year} {2001})},\ \Eprint
  {https://arxiv.org/abs/cond-mat/0006056} {cond-mat/0006056} \BibitemShut
  {NoStop}%
\bibitem [{\citenamefont {Doussal}\ \emph {et~al.}(2002)\citenamefont
  {Doussal}, \citenamefont {Wiese},\ and\ \citenamefont
  {Chauve}}]{LeDoussalWieseChauve2002}%
  \BibitemOpen
  \bibfield  {author} {\bibinfo {author} {\bibfnamefont {P.~L.}\ \bibnamefont
  {Doussal}}, \bibinfo {author} {\bibfnamefont {K.}~\bibnamefont {Wiese}},\
  and\ \bibinfo {author} {\bibfnamefont {P.}~\bibnamefont {Chauve}},\
  }\bibfield  {title} {\bibinfo {title} {2-loop functional renormalization
  group analysis of the depinning transition},\ }\href
  {https://doi.org/10.1103/PhysRevB.66.174201} {\bibfield  {journal} {\bibinfo
  {journal} {Phys. Rev. B}\ }\textbf {\bibinfo {volume} {66}},\ \bibinfo
  {pages} {174201} (\bibinfo {year} {2002})},\ \Eprint
  {https://arxiv.org/abs/cond-mat/0205108} {cond-mat/0205108} \BibitemShut
  {NoStop}%
\bibitem [{\citenamefont {Doussal}\ \emph {et~al.}(2004)\citenamefont
  {Doussal}, \citenamefont {Wiese},\ and\ \citenamefont
  {Chauve}}]{LeDoussalWieseChauve2003}%
  \BibitemOpen
  \bibfield  {author} {\bibinfo {author} {\bibfnamefont {P.~L.}\ \bibnamefont
  {Doussal}}, \bibinfo {author} {\bibfnamefont {K.}~\bibnamefont {Wiese}},\
  and\ \bibinfo {author} {\bibfnamefont {P.}~\bibnamefont {Chauve}},\
  }\bibfield  {title} {\bibinfo {title} {Functional renormalization group and
  the field theory of disordered elastic systems},\ }\href
  {https://doi.org/10.1103/PhysRevE.69.026112} {\bibfield  {journal} {\bibinfo
  {journal} {Phys. Rev. E}\ }\textbf {\bibinfo {volume} {69}},\ \bibinfo
  {pages} {026112} (\bibinfo {year} {2004})},\ \Eprint
  {https://arxiv.org/abs/cond-mat/0304614} {cond-mat/0304614} \BibitemShut
  {NoStop}%
\bibitem{footnote2}
{We show below that the   signal   specified in \Eq{Delta-def} is the same as that obtained by supposing a random force.}
%
\bibitem{footnote-stiffness-trap}
{The stiffness per trap is about $250 \rm pN/\mu m$  \cite{BercyThesis}, leading to half this value for the two traps. At the plateau start, the strands reduce this to  $m^{2} = 55\pm 5 \rm pN/\mu m$, see Fig.\ \ref{1}.}
%
\bibitem [{\citenamefont {Doussal}\ and\ \citenamefont
  {Wiese}(2009)}]{LeDoussalWiese2008a}%
  \BibitemOpen
  \bibfield  {author} {\bibinfo {author} {\bibfnamefont {P.~L.}\ \bibnamefont
  {Doussal}}\ and\ \bibinfo {author} {\bibfnamefont {K.}~\bibnamefont
  {Wiese}},\ }\bibfield  {title} {\bibinfo {title} {Driven particle in a random
  landscape: disorder correlator, avalanche distribution and extreme value
  statistics of records},\ }\href {https://doi.org/10.1103/PhysRevE.79.051105}
  {\bibfield  {journal} {\bibinfo  {journal} {Phys. Rev. E}\ }\textbf {\bibinfo
  {volume} {79}},\ \bibinfo {pages} {051105} (\bibinfo {year} {2009})},\
  \Eprint {https://arxiv.org/abs/arXiv:0808.3217} {arXiv:0808.3217}
  \BibitemShut {NoStop}%
%
\bibitem [{\citenamefont {{Le Doussal}}(2006)}]{LeDoussal2006b}%
  \BibitemOpen
  \bibfield  {author} {\bibinfo {author} {\bibfnamefont {P.}~\bibnamefont {{Le
  Doussal}}},\ }\bibfield  {title} {\bibinfo {title} {Finite temperature
  {Functional RG}, droplets and decaying {Burgers} turbulence},\ }\href
  {https://doi.org/10.1209/epl/i2006-10295-1} {\bibfield  {journal} {\bibinfo
  {journal} {Europhys. Lett.}\ }\textbf {\bibinfo {volume} {76}},\ \bibinfo
  {pages} {457} (\bibinfo {year} {2006})},\ \Eprint
  {https://arxiv.org/abs/cond-mat/0605490} {cond-mat/0605490} \BibitemShut
  {NoStop}%
\bibitem [{\citenamefont {{Le Doussal}}\ and\ \citenamefont
  {Wiese}(2007)}]{LeDoussalWiese2006a}%
  \BibitemOpen
  \bibfield  {author} {\bibinfo {author} {\bibfnamefont {P.}~\bibnamefont {{Le
  Doussal}}}\ and\ \bibinfo {author} {\bibfnamefont {K.}~\bibnamefont
  {Wiese}},\ }\bibfield  {title} {\bibinfo {title} {How to measure {Functional
  RG} fixed-point functions for dynamics and at depinning},\ }\href
  {https://doi.org/10.1209/0295-5075/77/66001} {\bibfield  {journal} {\bibinfo
  {journal} {EPL}\ }\textbf {\bibinfo {volume} {77}},\ \bibinfo {pages} {66001}
  (\bibinfo {year} {2007})},\ \Eprint {https://arxiv.org/abs/cond-mat/0610525}
  {cond-mat/0610525} \BibitemShut {NoStop}%
%
\bibitem{WieseLeDoussal2006}
K.J. Wiese and P.~Le Doussal,
\newblock {Functional renormalization for disordered systems: Basic recipes
  and gourmet dishes},
\newblock \href{http://math-mprf.org/journal/articles/id1143/}{Markov Processes Relat. Fields {\bf 13}, 777--818 (2007)},
\newblock \arxiv{cond-mat/0611346}.
%
\bibitem [{\citenamefont {Zinn-Justin}(1989)}]{Zinn}%
  \BibitemOpen
  \bibfield  {author} {\bibinfo {author} {\bibfnamefont {J.}~\bibnamefont
  {Zinn-Justin}},\ }\href
  {https://doi.org/10.1093/acprof:oso/9780199227198.001.0001} {\emph {\bibinfo
  {title} {Quantum Field Theory and Critical Phenomena}}}\ (\bibinfo
  {publisher} {Oxford University Press},\ \bibinfo {address} {Oxford},\
  \bibinfo {year} {1989})\BibitemShut {NoStop}%
\bibitem [{\citenamefont {Weinberg}(1995)}]{WeinbergBook123}%
  \BibitemOpen
  \bibfield  {author} {\bibinfo {author} {\bibfnamefont {S.}~\bibnamefont
  {Weinberg}},\ }\href {https://doi.org/10.1017/CBO9781139644167} {\emph
  {\bibinfo {title} {The Quantum Theory of Fields}}},\ Vol.\ \bibinfo {volume}
  {1-3}\ (\bibinfo  {publisher} {Cambridge University Press},\ \bibinfo {year}
  {1995})\BibitemShut {NoStop}%
\bibitem [{\citenamefont {Domb}\ \emph {et~al.}(2001)\citenamefont {Domb},
  \citenamefont {Green},\ and\ \citenamefont
  {Lebowitz}}]{PhaseTransitionsAndCriticalPhenomenaSeries}%
  \BibitemOpen
  \bibinfo {editor} {\bibfnamefont {C.}~\bibnamefont {Domb}}, \bibinfo {editor}
  {\bibfnamefont {M.}~\bibnamefont {Green}},\ and\ \bibinfo {editor}
  {\bibfnamefont {J.}~\bibnamefont {Lebowitz}},\ eds.,\ \href@noop {} {\emph
  {\bibinfo {title}
  {\href{https://en.wikipedia.org/wiki/Phase_Transitions_and_Critical_Phenomena}{Phase
  Transitions and Critical Phenomena}}}},\ Vol.\ \bibinfo {volume} {1-19}\
  (\bibinfo  {publisher} {Academic Press},\ \bibinfo {address} {London},\
  \bibinfo {year} {1972-2001})\BibitemShut {NoStop}%
\bibitem [{\citenamefont {Middleton}\ \emph {et~al.}(2007)\citenamefont
  {Middleton}, \citenamefont {{Le~Doussal}},\ and\ \citenamefont
  {Wiese}}]{MiddletonLeDoussalWiese2006}%
  \BibitemOpen
  \bibfield  {author} {\bibinfo {author} {\bibfnamefont {A.}~\bibnamefont
  {Middleton}}, \bibinfo {author} {\bibfnamefont {P.}~\bibnamefont
  {{Le~Doussal}}},\ and\ \bibinfo {author} {\bibfnamefont {K.}~\bibnamefont
  {Wiese}},\ }\bibfield  {title} {\bibinfo {title} {Measuring functional
  renormalization group fixed-point functions for pinned manifolds},\ }\href
  {https://doi.org/10.1103/PhysRevLett.98.155701} {\bibfield  {journal}
  {\bibinfo  {journal} {Phys. Rev. Lett.}\ }\textbf {\bibinfo {volume} {98}},\
  \bibinfo {pages} {155701} (\bibinfo {year} {2007})},\ \Eprint
  {https://arxiv.org/abs/cond-mat/0606160} {cond-mat/0606160} \BibitemShut
  {NoStop}%
\bibitem [{\citenamefont {Rosso}\ \emph {et~al.}(2007)\citenamefont {Rosso},
  \citenamefont {{Le~Doussal}},\ and\ \citenamefont
  {Wiese}}]{RossoLeDoussalWiese2006a}%
  \BibitemOpen
  \bibfield  {author} {\bibinfo {author} {\bibfnamefont {A.}~\bibnamefont
  {Rosso}}, \bibinfo {author} {\bibfnamefont {P.}~\bibnamefont
  {{Le~Doussal}}},\ and\ \bibinfo {author} {\bibfnamefont {K.}~\bibnamefont
  {Wiese}},\ }\bibfield  {title} {\bibinfo {title} {Numerical calculation of
  the functional renormalization group fixed-point functions at the depinning
  transition},\ }\href {https://doi.org/10.1103/PhysRevB.75.220201} {\bibfield
  {journal} {\bibinfo  {journal} {Phys. Rev. B}\ }\textbf {\bibinfo {volume}
  {75}},\ \bibinfo {pages} {220201} (\bibinfo {year} {2007})},\ \Eprint
  {https://arxiv.org/abs/cond-mat/0610821} {cond-mat/0610821} \BibitemShut
  {NoStop}%
%
\bibitem{note-noisy-data}
{For the   noisy data at hand, this procedure is more stable than the one used in Refs.~\cite{MiddletonLeDoussalWiese2006,RossoLeDoussalWiese2006a,LeDoussalWieseMoulinetRolley2009}, where $\Delta(w)$ was rescaled to have integral 1.}
%
\bibitem [{\citenamefont {{Le~Doussal}}\ and\ \citenamefont
  {Wiese}(2009)}]{LeDoussalWiese2008c}%
  \BibitemOpen
  \bibfield  {author} {\bibinfo {author} {\bibfnamefont {P.}~\bibnamefont
  {{Le~Doussal}}}\ and\ \bibinfo {author} {\bibfnamefont {K.}~\bibnamefont
  {Wiese}},\ }\bibfield  {title} {\bibinfo {title} {Size distributions of
  shocks and static avalanches from the functional renormalization group},\
  }\href {https://doi.org/10.1103/PhysRevE.79.051106} {\bibfield  {journal}
  {\bibinfo  {journal} {Phys. Rev. E}\ }\textbf {\bibinfo {volume} {79}},\
  \bibinfo {pages} {051106} (\bibinfo {year} {2009})},\ \Eprint
  {https://arxiv.org/abs/arXiv:0812.1893} {arXiv:0812.1893} \BibitemShut
  {NoStop}%
\bibitem [{\citenamefont {Breslauer}\ \emph {et~al.}(1986)\citenamefont
  {Breslauer}, \citenamefont {Frank}, \citenamefont {Bl\"ocker},\ and\
  \citenamefont {Marky}}]{BreslauerFrankBlockerMarky1986}%
  \BibitemOpen
  \bibfield  {author} {\bibinfo {author} {\bibfnamefont {K.~J.}\ \bibnamefont
  {Breslauer}}, \bibinfo {author} {\bibfnamefont {R.}~\bibnamefont {Frank}},
  \bibinfo {author} {\bibfnamefont {H.}~\bibnamefont {Bl\"ocker}},\ and\
  \bibinfo {author} {\bibfnamefont {L.~A.}\ \bibnamefont {Marky}},\ }\bibfield
  {title} {\bibinfo {title} {Predicting {DNA} duplex stability from the base
  sequence},\ }\href {https://doi.org/10.1073/pnas.83.11.3746} {\bibfield
  {journal} {\bibinfo  {journal} {PNAS}\ }\textbf {\bibinfo {volume} {83}},\
  \bibinfo {pages} {3746} (\bibinfo {year} {1986})}\BibitemShut {NoStop}%
\bibitem [{\citenamefont {Sugimoto}\ \emph {et~al.}(1995)\citenamefont
  {Sugimoto}, \citenamefont {Nakano}, \citenamefont {Katoh}, \citenamefont
  {Matsumura}, \citenamefont {Nakamuta}, \citenamefont {Ohmichi}, \citenamefont
  {Yoneyama},\ and\ \citenamefont
  {Sasaki}}]{SugimotoNakanoKatohMatsumuraNakamutaOhmichiYoneyamaSasaki1995}%
  \BibitemOpen
  \bibfield  {author} {\bibinfo {author} {\bibfnamefont {N.}~\bibnamefont
  {Sugimoto}}, \bibinfo {author} {\bibfnamefont {S.}~\bibnamefont {Nakano}},
  \bibinfo {author} {\bibfnamefont {M.}~\bibnamefont {Katoh}}, \bibinfo
  {author} {\bibfnamefont {A.}~\bibnamefont {Matsumura}}, \bibinfo {author}
  {\bibfnamefont {H.}~\bibnamefont {Nakamuta}}, \bibinfo {author}
  {\bibfnamefont {T.}~\bibnamefont {Ohmichi}}, \bibinfo {author} {\bibfnamefont
  {M.}~\bibnamefont {Yoneyama}},\ and\ \bibinfo {author} {\bibfnamefont
  {M.}~\bibnamefont {Sasaki}},\ }\bibfield  {title} {\bibinfo {title}
  {Thermodynamic parameters to predict stability of {RNA/DNA} hybrid
  duplexes},\ }\bibfield  {booktitle} {\emph {\bibinfo {booktitle}
  {Biochemistry}},\ }\href {https://doi.org/10.1021/bi00035a029} {\bibfield
  {journal} {\bibinfo  {journal} {Biochemistry}\ }\textbf {\bibinfo {volume}
  {34}},\ \bibinfo {pages} {11211} (\bibinfo {year} {1995})}\BibitemShut
  {NoStop}%
%
\bibitem{note-F-w}
{In Ref.~\cite{LeDoussalWiese2008c} drops  in position $u$ of size $S$ are considered, related via \Eq{1} to    force drops  as $\delta F = m^2 S$.}
%
\bibitem [{\citenamefont {Chauve}\ \emph {et~al.}(2000)\citenamefont {Chauve},
  \citenamefont {Giamarchi},\ and\ \citenamefont
  {Doussal}}]{ChauveGiamarchiLeDoussal2000}%
  \BibitemOpen
  \bibfield  {author} {\bibinfo {author} {\bibfnamefont {P.}~\bibnamefont
  {Chauve}}, \bibinfo {author} {\bibfnamefont {T.}~\bibnamefont {Giamarchi}},\
  and\ \bibinfo {author} {\bibfnamefont {P.~L.}\ \bibnamefont {Doussal}},\
  }\bibfield  {title} {\bibinfo {title} {Creep and depinning in disordered
  media},\ }\href {https://doi.org/10.1103/PhysRevB.62.6241} {\bibfield
  {journal} {\bibinfo  {journal} {Phys. Rev. B}\ }\textbf {\bibinfo {volume}
  {62}},\ \bibinfo {pages} {6241} (\bibinfo {year} {2000})},\ \Eprint
  {https://arxiv.org/abs/cond-mat/0002299} {cond-mat/0002299} \BibitemShut
  {NoStop}%
\bibitem [{\citenamefont {Huguet}\ \emph {et~al.}(2009)\citenamefont {Huguet},
  \citenamefont {Forns},\ and\ \citenamefont {Ritort}}]{HuguetFornsRitort2009}%
  \BibitemOpen
  \bibfield  {author} {\bibinfo {author} {\bibfnamefont {J.}~\bibnamefont
  {Huguet}}, \bibinfo {author} {\bibfnamefont {N.}~\bibnamefont {Forns}},\ and\
  \bibinfo {author} {\bibfnamefont {F.}~\bibnamefont {Ritort}},\ }\bibfield
  {title} {\bibinfo {title} {Statistical properties of metastable intermediates
  in {DNA} unzipping},\ }\href {https://doi.org/10.1103/PhysRevLett.103.248106}
  {\bibfield  {journal} {\bibinfo  {journal} {Phys. Rev. Lett.}\ }\textbf
  {\bibinfo {volume} {103}},\ \bibinfo {pages} {248106} (\bibinfo {year}
  {2009})}\BibitemShut {NoStop}%
\bibitem{last-note}
{Note that $m^2$ here equals $k$ in Ref.~\cite{HuguetFornsRitort2009}, and $\rho_m$ here equals $n_c$ there. The theory for the exponent $4/3$ is given in Ref.~\cite{LeDoussalWieseChauve2003}, Eq.~(4.22), setting $d=0$, i.e.\ $\epsilon=4$ there.} 
\end{thebibliography}

%

\end{document}